# Discovering physical mechanisms from experiment–simulation mismatches

Yue Li[1], Penghui Yang[2], Yushan Xiao[1], Zhonghan Zhang[1], Jianguo Huang[2], Yuhao Lu[2], Cuntai Guan[2], Bo An[2], Bijun Tang[1,*], Zheng Liu[1,3,*]

[1]School of Materials Science and Engineering, Nanyang Technological University, 50 Nanyang Avenue, Singapore 639798, Singapore.

[2]College of Computing and Data Science, Nanyang Technological University, Singapore 639798, Singapore.

[3]Institute for Functional Intelligent Materials, National University of Singapore, Singapore, Singapore.

*Correspondence: bjtang@ntu.edu.sg; z.liu@ntu.edu.sg

## Abstract

Scientific discovery often begins where observation and prediction disagree. As computation and machine learning survey chemical space, experiment–simulation mismatches are exposed at scale, while tracing them to physical mechanisms remains expert-led. Here we present eXplainable DFT (XDFT), a self-evolving agent that turns this process into an executable search. XDFT formalizes candidate mechanisms as executable hypotheses, adjudicates their consequences against experiment and distils trajectories into priors for later searches. This couples a solving loop from mismatch to mechanism with a learning loop through which solving changes the solver. Across 112 source-audited cases in which standard calculations predict a metal whereas experiments find a semiconductor, XDFT resolved 105 with evidence-supported mechanisms within a single-GPU envelope. After 60 cases, the resolving mechanism ranked among the first three hypotheses for 80% of held-out cases, up from 7% under the initial expert prior. XDFT also returned evidence-graded mechanisms for seven expert-curated questions about physical mechanisms. These results establish experiment–simulation mismatches as tractable starting points for scientific agents that discover physical mechanisms while learning how to find the next.

## Introduction

Some of the most consequential advances in science have begun with reproducible observations that prevailing theories could not explain. Mercury's anomalous perihelion precession exposed a limit of Newtonian gravitation and was later accounted for by general relativity.[1] The failure of classical physics to reproduce the black-body spectrum led Planck to introduce energy quanta.[2] In both cases, reproducible disagreement marked the boundary of the prevailing description and made the missing mechanism the scientific target.

Materials science is now producing such opportunities at rapidly expanding scale. Open databases and high-throughput platforms such as the Materials Project made first-principles properties searchable across chemical space;[3] GNoME extended stable-structure searches into the millions;[4] MatterGen enabled inverse materials design;[5] and universal interatomic potentials including M3GNet, CHGNet and MACE extended DFT-trained simulation to larger chemical, length and time scales.[6–8] Together, these advances have transformed the reach of computational materials science. Their power comes from applying standardized physical descriptions broadly or learning recurring relations from data-rich regions. Anomalies expose degrees of freedom omitted from the calculation or scarce in the training data. Experiment–calculation mismatches can therefore contain electron correlation, hidden structures, magnetic order, defects or relativistic interactions outside the scope of general-purpose models. Resolving them requires experimental and computational experts to turn contextual knowledge into competing mechanisms, construct non-standard calculations and judge their consequences against observation. The resulting challenge is a general route from mismatch to mechanism.

Scientific agents are beginning to connect generative reasoning with scientific software, allowing hypotheses to be tested rather than merely stated.[9–11] Here we present eXplainable DFT (XDFT), a self-evolving agent that searches from experiment–simulation mismatch to physical mechanism. XDFT formalizes qualitative explanations as executable hypotheses, adjudicates their consequences through first-principles calculations and returns each outcome to the search as prior knowledge. We screened more than 100,000 band-gap records from four literature and data sources,[12–15] cross-compared them with standard Materials Project calculations and traced the resulting candidates to primary publications. This yielded 112 source-audited mismatches in which semilocal PBE/PBE+U calculations predict a metal but experiment reports a semiconductor.[3,16] XDFT resolved 105 within a single-GPU envelope. Figure 1 defines the anomaly population and tests recovery of the observed electronic state. Its formalization, adjudication and prior update couple the solving of each case to learning from it (Fig. 2).

Solving also changes the solver. Accepted and rejected hypotheses update statistical and semantic priors, moving successful mechanisms towards the front of later searches and reopening cases that the earlier policy could not resolve (Fig. 3). We then generalized the same architecture beyond a fixed band-gap discrepancy by representing seven expert-curated questions as mismatches between observation and expectation. XDFT returned an evidence-graded mechanism for each, including a case that remains contested in the literature (Fig. 4 and Supplementary Information). Thus, mechanism is the endpoint of each case, while its adjudicated trajectory becomes knowledge for finding the next.

## Results and discussion

### A controlled anomaly benchmark separates prediction from mechanism search

To establish mismatch-to-mechanism reasoning across a population, we selected a deliberately bounded materials problem. Semiconductor–metal mismatches provide a data-rich and executable testbed: standard Materials Project PBE/PBE+$U$ calculations predict a metal, primary experiments report a semiconductor, and any proposed mechanism must account for the missing gap. We screened more than 100,000 band-gap records from four literature and data sources[12–15] and cross-compared them with the standard calculations. Composition-level deduplication yielded 613 candidates; tracing each to its primary publication retained 112 source-audited anomalies, divided into an 87-case evolving set and an independent 25-case probe.

These anomalies lie where models optimized for average predictive performance lose reliability. Sixty-six evolving cases occur in the 4,604-composition Matbench experimental-gap set.[12,17] They constitute only 1.4% of that space, yet span its major anion families and form no distinct region in composition space (Fig. 1a). Composition-only predictors trained against experimental gaps (Darwin, CrabNet and MODNet)[18–20] report general-benchmark mean absolute errors of 0.287–0.333 eV, which rise to 0.932–0.980 eV on the anomalies. Seven structure-based graph neural networks trained to reproduce PBE gaps, including CGCNN and ALIGNN,[21,22] reach 0.156–0.297 eV on their general benchmark, yet deviate from the source-audited experimental gaps by 1.226–1.382 eV on the anomaly set (Supplementary Information). Because the two model classes use different training targets, these values are not a common ranking; repeating the comparisons with matched reference labels preserves the same qualitative collapse (Supplementary Information). Strong average-distribution performance therefore does not transfer to this anomaly tail, motivating a search for physical mechanisms beyond property interpolation.

We next asked whether each system could recover the observed state of the 25 held-out anomalies. The best composition-only and structure-based models identified 19 and 12 experimental semiconductors, respectively; five frontier language models identified 22–25, and XDFT identified 24 (Fig. 1b; symmetric control in Supplementary Information). Recovering the observed state is therefore only the entry condition. The scientific task is to identify the mechanism that makes the standard calculation fail and to support it with executable evidence.

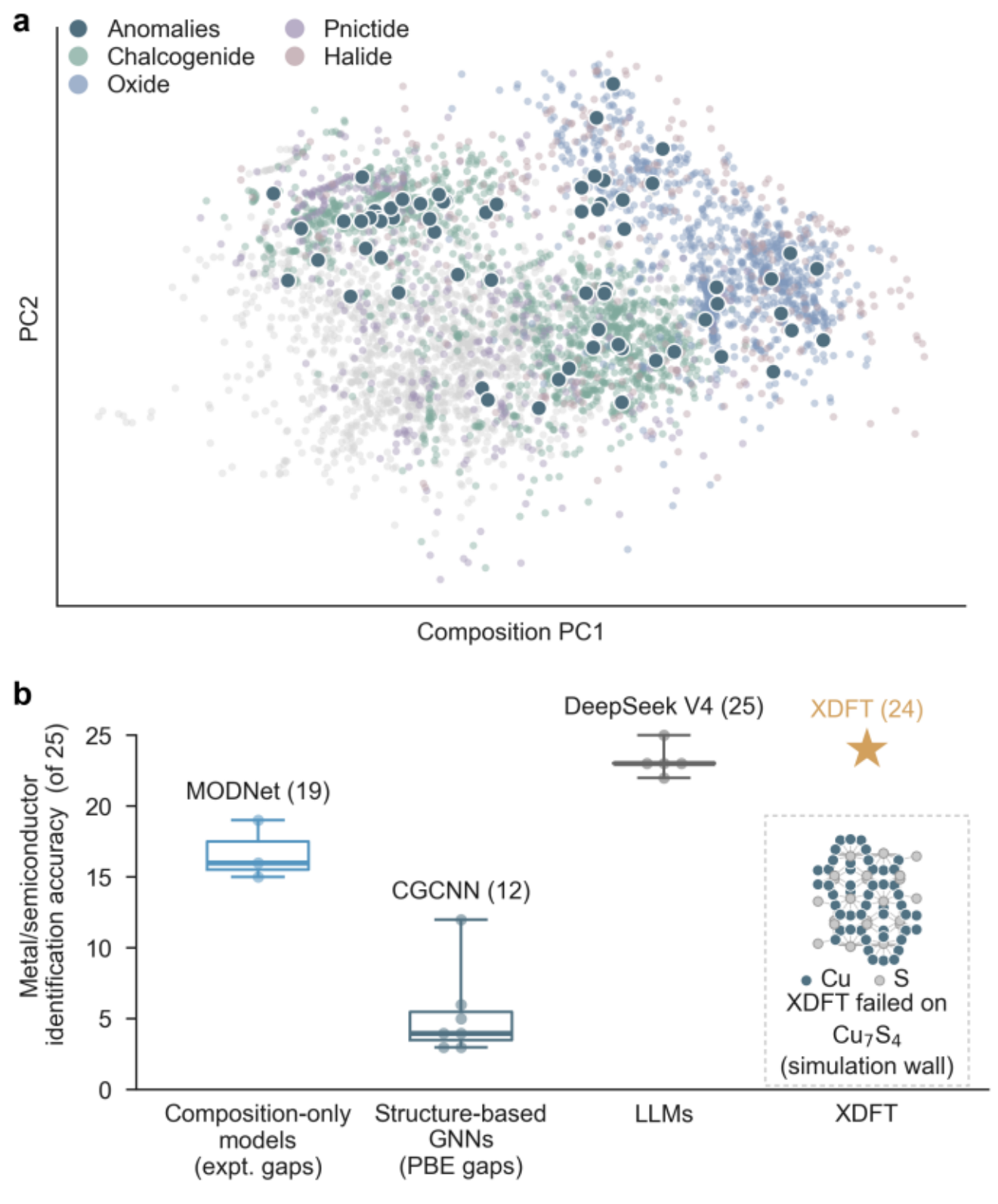

**Fig. 1 | A controlled semiconductor–metal anomaly benchmark.** a, Principal-component projection of 4,604 experimental-gap compositions; dark points mark 66 source-audited mismatches. b, Recovery of the experimental semiconductor state in the 25-case probe by composition-only models (n = 3), structure-based graph neural networks (n = 7), frontier language models (n = 5) and XDFT. Inset, $Cu_7S_4$, the only probe unresolved within the single-GPU envelope. A symmetric metal control is reported in the Supplementary Information.

## Formalization and adjudication turn mechanisms into executable tests

Efficient mechanism search requires each costly calculation to eliminate alternatives and determine the next physical test. XDFT achieves this through three linked operations: formalization converts an explanation into an executable intervention; adjudication compares its calculated consequence with experiment; and prior update carries the outcome into subsequent choices (Fig. 2). Together, these operations couple a solving loop for the present mismatch to a learning loop that improves later searches.

The proposal half of each round turns a physical idea into a specific calculation. The Reasoner receives the mismatch together with the current prior, retrieves the relevant structure[3,24] and selects a hypothesis card. Each card binds a human-readable physical claim to an executable action. Together with the registered observable and acceptance criterion used in adjudication, it defines a discriminating test. The Manipulator then compiles the action into a concrete intervention, such as a modified structure or electronic-structure calculation. Formalization thereby turns a qualitative explanation into a searchable object.

The evidence half converts that calculation into a decision. The Simulator executes the compiled first-principles calculation with VASP[23], and the Comparator adjudicates the result against the registered observation and competing explanations. A rejected hypothesis returns to the Reasoner together with its calculated outcome, constraining the next selection. A hypothesis that passes adjudication closes the solving loop, and its supported physical claim becomes the mechanism. The resulting mechanism is supported under the registered observable, structure, physical approximation, numerical settings and criterion; additional evidence can reopen the case.

The same decision also drives learning (Fig. 2). Outcomes update a statistical prior over successful cards, whereas solved trajectories are distilled into semantic rules. When no existing card captures a viable explanation, XDFT generates and vets a new one. These priors re-enter the Reasoner, coupling progress on the present case to the starting point of the next. Self-evolution should therefore produce two measurable gains: resolving mechanisms should move towards the front of later searches, and the evolved policy should reach cases left unresolved by its earlier state. Figure 3 quantifies both effects.

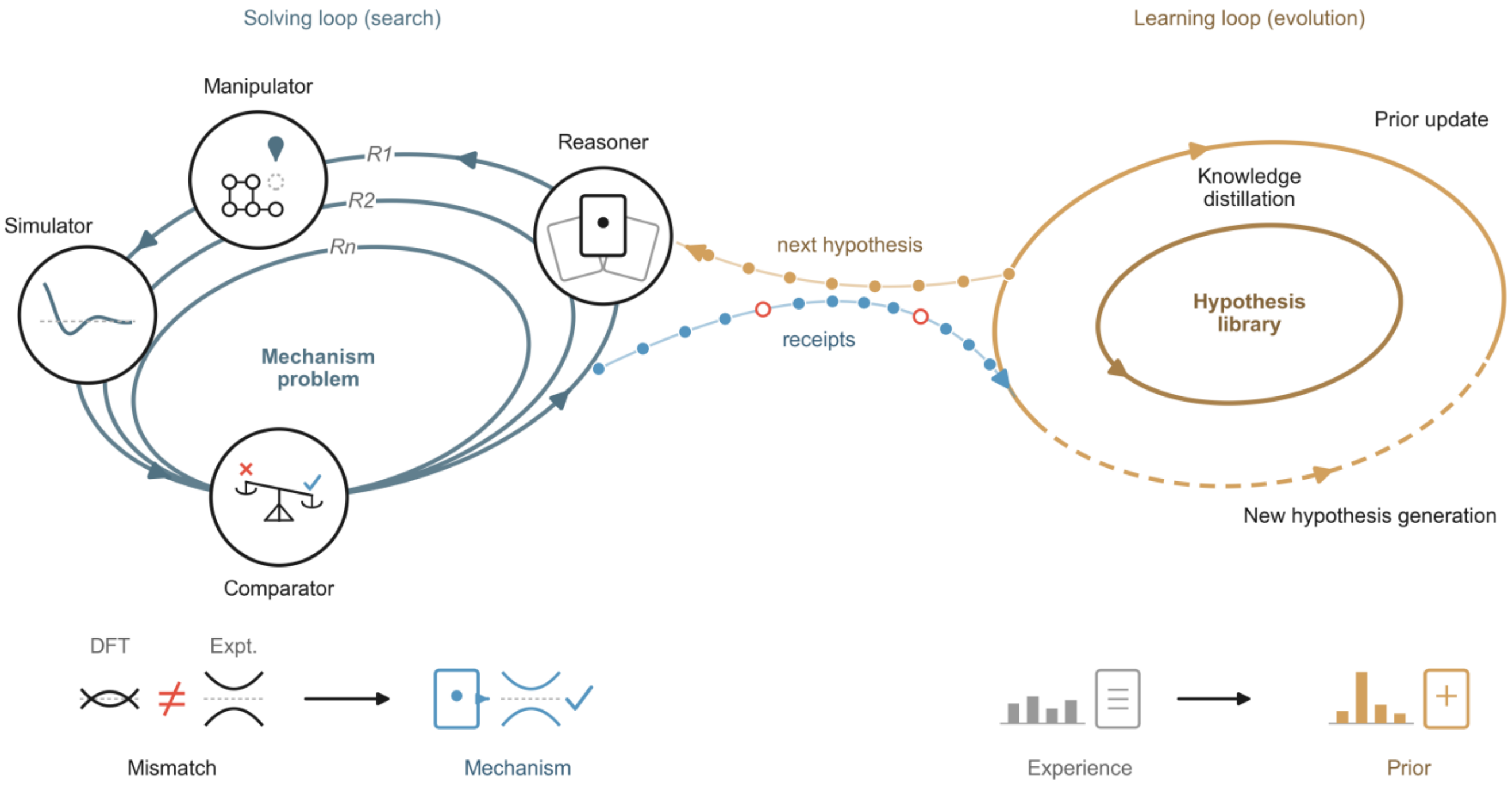

**Fig. 2 | Formalization, adjudication and prior update couple solving to learning.** XDFT searches from mismatch to mechanism through a Reasoner, Manipulator, Simulator and Comparator, while returning adjudicated outcomes as statistical and semantic priors for subsequent searches.

## Adjudicated experience evolves mechanism search

We tested whether adjudicated experience changed later searches in two ways: whether the resolving card moved forward in the ranking, and whether the evolved policy recovered cases left unresolved by its earlier state (Fig. 3a). The 87 searches generated 345 adjudicated selections that updated a statistical prior and yielded semantic rules for subsequent choices.

The statistical channel measurably changes mechanism ordering. On a 15-material completion-order hold-out, the fraction whose eventual resolving card lies within the first three positions rises from 7% under the expert prior to 80% after 60 cases, compared with about 9% for random ordering (Fig. 3b). The resolved trajectories also reveal chemistry-dependent structure in the experience from which the prior is learned (Fig. 3c): magnetic-order and Hubbard-$U$ cards dominate resolved d- and f-block cases, whereas polymorph reselection dominates main-group cases.

Semantic distillation captures when successful searches must change physical direction. Seven of the 69 resolved materials switch between method and structure cards, and replaying the distilled rules at those decisions supports six of the seven moves (Fig. 3d). PdO makes such a redirection concrete: four Hubbard-U calculations remain below the gap criterion before polymorph reselection opens it (Fig. 3e). This replay shows that the distilled rules encode successful cross-family redirections, although it does not isolate their causal contribution from statistical exhaustion.

The evolved loop also expands the reach of mechanism search. Applied to the 18 cases left unresolved earlier in the campaign, it resolves 12; the remaining six terminate at the configured single-GPU simulation wall (Fig. 3f). Together, these results locate evolution in the search policy: the ordering of mechanism tests, rules that encode successful cross-family redirections and the set of cases resolved within the computational envelope. The benchmark establishes these gains within one anomaly class with a fixed numerical comparator. Extending the same search to open scientific questions requires a generalized mismatch that formalizes the conflicting observation, expectation and case-specific evidence. Figure 4 implements this extension.

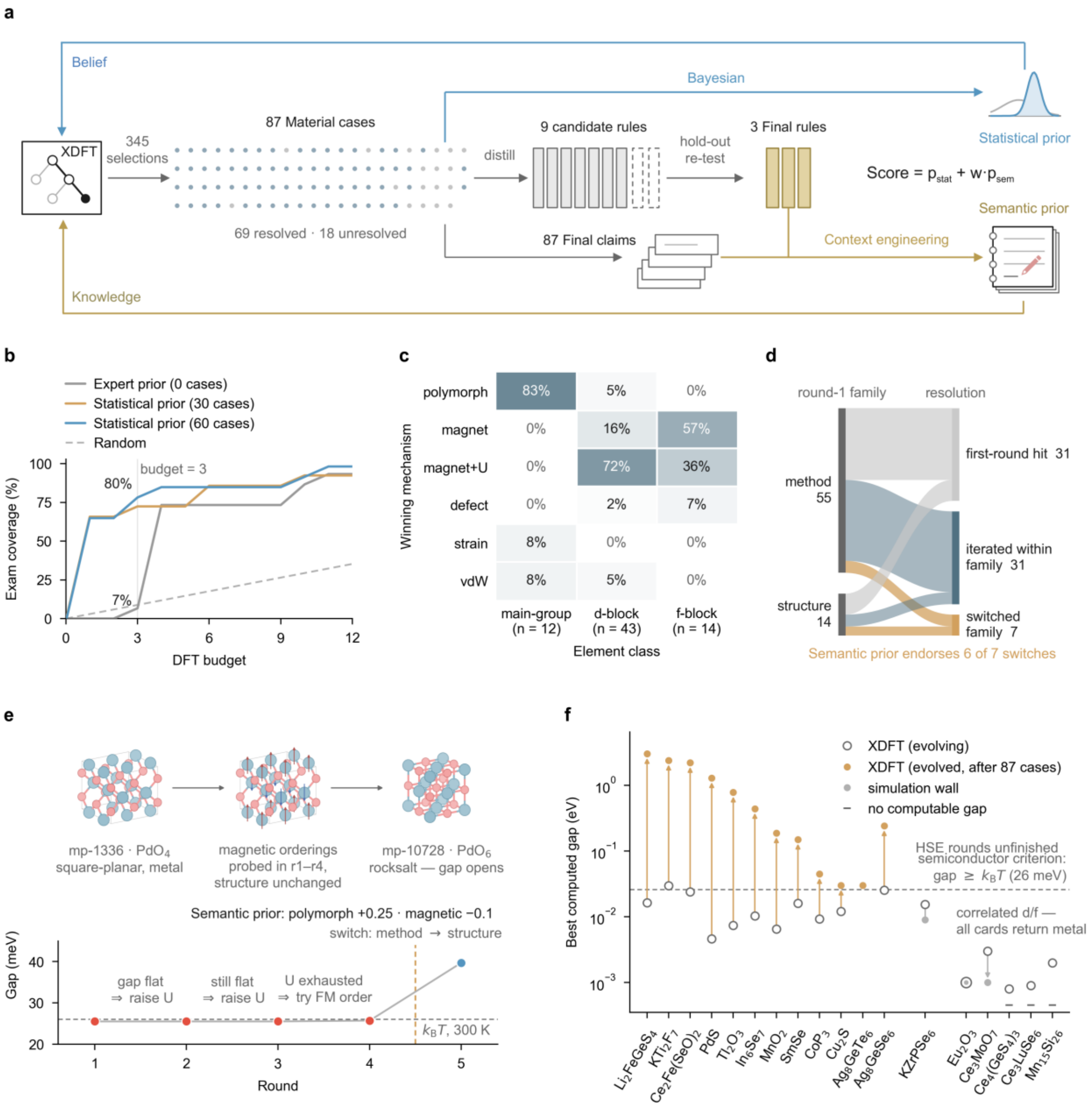

**Fig. 3 | Adjudicated experience changes the ordering and reach of mechanism search.** a, Outcomes from 87 cases and 345 selections update statistical and semantic priors. b, Top-k coverage of the resolving card under the expert prior and priors updated after 30 or 60 cases; dashed, random ordering. At k = 3, coverage rises from 7% to 80%. c, Winning mechanism family by element class for 69 resolved cases. d, Search paths within and between card families; the semantic prior supports six of seven switches. e, PdO trajectory from four Hubbard-U tests to a gap-opening rocksalt polymorph; dashed, kBT = 26 meV. f, Best gaps for 18 previously unresolved materials before and after evolution. Twelve cross the criterion; six stop at the simulation wall.

## Generalized mismatches make open questions of mechanism executable

The benchmark so far fixes both the anomaly and the adjudication criterion: experiment reports a semiconductor where the standard calculation gives a metal. We next asked whether the same architecture could investigate less structured questions of physical mechanism. XDFT represents each question as a generalized mismatch between an observation and an expectation, together with the baseline calculation, discriminating observables and evidence standard needed to evaluate candidate mechanisms (Fig. 4a). It retains the evolved prior, rules, cards and search loop from the benchmark setting. We evaluated seven expert-curated questions, three here and four in the Supplementary Information; six have established published mechanisms and $WTe_2$ preserves a disagreement in the literature.

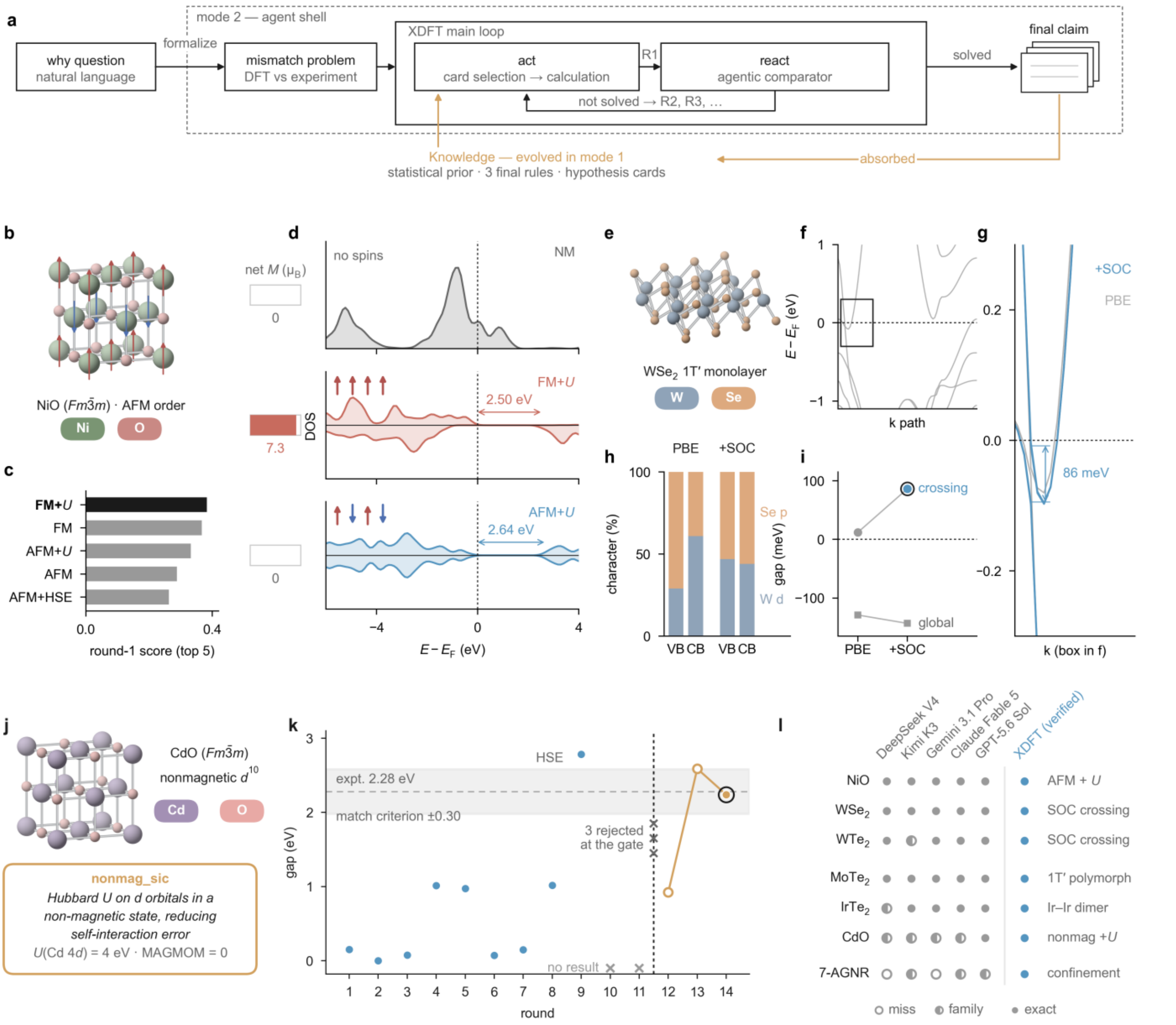

**Fig. 4 | Generalized mismatches make open questions of mechanism executable.** a, A question is formalized as a mismatch between observation and expectation, then searched through calculation and adjudication. b–d, NiO magnetic controls support antiferromagnetic order with on-site Coulomb repulsion. e–i, In 1T′-$WSe_2$, spin–orbit coupling increases a local band separation from 12 to 86 meV while the global state remains semimetallic. j,k, In CdO, generation adds a new executable intervention after eleven library mechanisms fail and reaches 2.24 eV. l, Evidence levels returned for seven expert-curated questions. Open, mechanism miss; half-filled, mechanism family; filled, exact literature mechanism; blue, computational adjudication.

NiO requires magnetic controls because several mechanisms reproduce a similar gap (Fig. 4b–d). The evolved prior ranks ferromagnetic order with Hubbard U (FM+U) first, and XDFT expands the test to non-magnetic (NM), FM+U and antiferromagnetic order with Hubbard U (AFM+U) calculations (Fig. 4c). NM remains metallic, whereas FM+U and AFM+U open similar gaps of 2.50 and 2.64 eV at near-equal local Ni moments. Their net cell moments, 7.3 and 0 μB, identify the antiferromagnetic solution (Fig. 4d), supporting a mechanism that combines antiferromagnetic order with on-site Coulomb repulsion.[25,26] The cubic cell adjudicates the magnetic class; resolving the type-II ground state of NiO requires a larger cell.[27]

For 1T′-$WSe_2$, several observables are needed to distinguish a local spin–orbit gap from a global insulating state (Fig. 4e–i). Without spin–orbit coupling, the crossing pair lies 12 meV apart and has distinct character, 71% Se p and 61% W d (Fig. 4f,h). Spin–orbit coupling increases the separation to 86 meV and mixes both branches to approximately equal W d and Se p character (Fig. 4g,h), while the global band overlap increases from 129 to 143 meV and the sheet remains semimetallic (Fig. 4i). The evidence supports a local spin–orbit gap at a crossing of distinct-character bands while preserving the semimetallic global state. The single-momentum projection establishes character mixing; the band-inversion assignment therefore remains attributable to the published prediction.[28]

CdO tests whether accumulated failures can expand the hypothesis space (Fig. 4j,k). Eleven predefined mechanisms fail, including a screened hybrid-functional calculation (HSE)[29] that overestimates the approximately 2.3-eV reference[30] by 0.51 eV. XDFT then generates and vets a new intervention that applies Hubbard U to Cd 4d states without spin polarization. The resulting 2.24-eV gap approaches the experimental reference and adds an executable action absent from the original library. Because the calculation yields an indirect Kohn–Sham gap whereas the 2.28-eV reference is probably direct, the mechanism remains conditional on the registered comparison criterion.

Across the seven questions, the outputs form an evidence ladder from a mechanism family, to a case-specific mechanism, to computational adjudication (Fig. 4l). Frontier language models often name the first two levels when the supplied record is diagnostic. The matrix records the forms of evidence returned by each system under a separate zero-shot protocol. XDFT returns an evidence-graded mechanism for all seven questions. Generalized mismatches thus expand the searched object from a scalar band-gap correction to an open question whose observation, expectation, discriminants and adjudication standard are explicitly registered. Seven curated questions establish this representational and operational extension; population-level performance remains to be measured.

## Conclusion

Experiment–simulation mismatches are commonly treated as prediction errors. XDFT turns them into mechanism-search problems by formalizing physical explanations, adjudicating them through first-principles interventions and converting completed searches into prior knowledge. Across a population-scale benchmark, this organization improved both the ordering of mechanism tests and the set of cases resolved within a fixed computational envelope; the same architecture also supported evidence-graded explanations for broader mechanistic questions. Self-evolution is therefore operational: solving one case changes how the next is approached.

The present study establishes this architecture within one anomaly class and seven curated extensions. Broader use will require domain-specific interventions, observables and adjudication criteria. The same structure can ultimately couple calculations with experiments that discriminate among competing explanations. In such a loop, every resolved mismatch yields two forms of knowledge: a physical mechanism for the present case and a prior that improves the search for the next.

## References

1 Einstein, A. Die Grundlage der allgemeinen Relativitätstheorie. *Ann. Phys.* **354**, 769–822, doi:10.1002/andp.19163540702 (1916).

2 Planck, M. Ueber das Gesetz der Energieverteilung im Normalspectrum. *Ann. Phys.* **309**, 553–563, doi:10.1002/andp.19013090310 (1901).

3 Jain, A. *et al.* Commentary: The Materials Project: a materials genome approach to accelerating materials innovation. *APL Mater.* **1**, 011002, doi:10.1063/1.4812323 (2013).

4 Merchant, A. *et al.* Scaling deep learning for materials discovery. *Nature* **624**, 80–85, doi:10.1038/s41586-023-06735-9 (2023).

5 Zeni, C. *et al.* A generative model for inorganic materials design. *Nature* **639**, 624–632, doi:10.1038/s41586-025-08628-5 (2025).

6 Chen, C. & Ong, S. P. A universal graph deep learning interatomic potential for the periodic table. *Nat. Comput. Sci.* **2**, 718–728, doi:10.1038/s43588-022-00349-3 (2022).

7 Deng, B. *et al.* CHGNet as a pretrained universal neural network potential for charge-informed atomistic modelling. *Nat. Mach. Intell.* **5**, 1031–1041, doi:10.1038/s42256-023-00716-3 (2023).

8 Batatia, I. *et al.* A foundation model for atomistic materials chemistry. *J. Chem. Phys.* **163**, 184110, doi:10.1063/5.0297006 (2025).

9 Romera-Paredes, B. *et al.* Mathematical discoveries from program search with large language models. *Nature* **625**, 468–475, doi:10.1038/s41586-023-06924-6 (2024).

10 Nägele, M. & Marquardt, F. Agentic exploration of physics models. *Phys. Rev. X* **16**, 031002, doi:10.1103/xnqc-q6nt (2026).

11 Liu, J. *et al.* VASPilot: MCP-facilitated multi-agent intelligence for autonomous VASP simulations. *Chin. Phys. B* **34**, 117106, doi:10.1088/1674-1056/ae0681 (2025).

12 Zhuo, Y., Mansouri Tehrani, A. & Brgoch, J. Predicting the band gaps of inorganic solids by machine learning. *J. Phys. Chem. Lett.* **9**, 1668–1673, doi:10.1021/acs.jpclett.8b00124 (2018).

13 Borlido, P. *et al.* Large-scale benchmark of exchange-correlation functionals for the determination of electronic band gaps of solids. *J. Chem. Theory Comput.* **15**, 5069–5079, doi:10.1021/acs.jctc.9b00322 (2019).

14 Madelung, O. *Semiconductors: Data Handbook*, 3rd edn, doi:10.1007/978-3-642-18865-7 (Springer, 2004).

15 Dong, Q. & Cole, J. M. Auto-generated database of semiconductor band gaps using ChemDataExtractor. *Sci. Data* **9**, 193, doi:10.1038/s41597-022-01294-6 (2022).

16 Perdew, J. P., Burke, K. & Ernzerhof, M. Generalized gradient approximation made simple. *Phys. Rev. Lett.* **77**, 3865–3868, doi:10.1103/PhysRevLett.77.3865 (1996).

17 Dunn, A., Wang, Q., Ganose, A., Dopp, D. & Jain, A. Benchmarking materials property prediction methods: the Matbench test set and Automatminer reference algorithm. *npj Comput. Mater.* **6**, 138, doi:10.1038/s41524-020-00406-3 (2020).

18 Xie, T. *et al.* Large language models as materials science adapted learners. Preprint at https://doi.org/10.21203/rs.3.rs-6752901/v1 (2025).

19 Wang, A. Y.-T., Kauwe, S. K., Murdock, R. J. & Sparks, T. D. Compositionally restricted attention-based network for materials property predictions. *npj Comput. Mater.* **7**, 77, doi:10.1038/s41524-021-00545-1 (2021).

20 De Breuck, P.-P., Hautier, G. & Rignanese, G.-M. Materials property prediction for limited datasets enabled by feature selection and joint learning with MODNet. *npj Comput. Mater.* **7**, 83, doi:10.1038/s41524-021-00552-2 (2021).

21 Xie, T. & Grossman, J. C. Crystal graph convolutional neural networks for an accurate and interpretable prediction of material properties. *Phys. Rev. Lett.* **120**, 145301, doi:10.1103/PhysRevLett.120.145301 (2018).

22 Choudhary, K. & DeCost, B. Atomistic line graph neural network for improved materials property predictions. *npj Comput. Mater.* **7**, 185, doi:10.1038/s41524-021-00650-1 (2021).

23 Kresse, G. & Furthmüller, J. Efficient iterative schemes for *ab initio* total-energy calculations using a plane-wave basis set. *Phys. Rev. B* **54**, 11169–11186, doi:10.1103/PhysRevB.54.11169 (1996).

24 Haastrup, S. *et al.* The Computational 2D Materials Database: high-throughput modeling and discovery of atomically thin crystals. *2D Mater.* **5**, 042002, doi:10.1088/2053-1583/aacfc1 (2018).

25 Sawatzky, G. A. & Allen, J. W. Magnitude and origin of the band gap in NiO. *Phys. Rev. Lett.* **53**, 2339–2342, doi:10.1103/PhysRevLett.53.2339 (1984).

26 Anisimov, V. I., Zaanen, J. & Andersen, O. K. Band theory and Mott insulators: Hubbard *U* instead of Stoner *I*. *Phys. Rev. B* **44**, 943–954, doi:10.1103/PhysRevB.44.943 (1991).

27 Roth, W. L. Magnetic structures of MnO, FeO, CoO, and NiO. *Phys. Rev.* **110**, 1333–1341, doi:10.1103/PhysRev.110.1333 (1958).

28 Qian, X., Liu, J., Fu, L. & Li, J. Quantum spin Hall effect in two-dimensional transition metal dichalcogenides. *Science* **346**, 1344–1347, doi:10.1126/science.1256815 (2014).

29 Heyd, J., Scuseria, G. E. & Ernzerhof, M. Hybrid functionals based on a screened Coulomb potential. *J. Chem. Phys.* **118**, 8207–8215, doi:10.1063/1.1564060 (2003).

30 Koffyberg, F. P. Thermoreflectance spectra of CdO: band gaps and band-population effects. *Phys. Rev. B* **13**, 4470–4476, doi:10.1103/PhysRevB.13.4470 (1976).